\documentclass[conference]{IEEEtran}
\IEEEoverridecommandlockouts

\usepackage{cite}

\usepackage{amsmath,amssymb,amsfonts}
\usepackage{algorithmic}
\usepackage{graphicx}
\usepackage{textcomp}
\usepackage{xcolor}
\usepackage{enumitem}
\usepackage{hyperref}

\usepackage[many]{tcolorbox}
\usepackage{wrapfig}
\usepackage{url}
\usepackage{float}
\usepackage{listings}
\usepackage[labelsep=colon,font=footnotesize]{caption}
\usepackage{xspace}

\DeclareFontFamily{U}{matha}{\hyphenchar\font45}
\DeclareFontShape{U}{matha}{m}{n}{
      <5> <6> <7> <8> <9> <10> gen * matha
      <10.95> matha10 <12> <14.4> <17.28> <20.74> <24.88> matha12
      }{}
\DeclareSymbolFont{matha}{U}{matha}{m}{n}

\def\tablename{Table} 
\renewcommand{\thetable}{\arabic{table}} 
\makeatletter
\patchcmd{\@makecaption}
  {\scshape}
  {}
  {}
  {}
\makeatletter
\patchcmd{\@makecaption}
  {\\}
  {:\ }
  {}
  {}
\makeatother

\newcommand{\codegen}{\textit{CodeGen1-Multi}\xspace}

\newcommand{\prom}{\texttt{prompt0}\xspace}
\newcommand{\promi}{\texttt{prompt1}\xspace}
\newcommand{\promii}{\texttt{prompt2}\xspace}
\newcommand{\promiii}{\texttt{prompt3}\xspace}
\newcommand{\promiv}{\texttt{prompt4}\xspace}

\newcommand{\EmergentEval}{{\textsc{EmergentEval}}\xspace}
\newcommand{\codeR}{{\textit{Bug Fixing}}\xspace}
\newcommand{\ct}{{\textit{Code Translation}}\xspace}
\newcommand{\cm}{{\textit{Commit Message Generation}}\xspace}

\newcommand{\exm}{\textit{EM}\xspace}
\newcommand{\bleu}{{\textit{BLEU}}\xspace}
\newcommand{\codebleu}{{\textit{CodeBLEU}}\xspace}
\newcounter{todoCount}

\newcounter{alejoCount}

\newcounter{questionCount}

\newtcolorbox{story}[1][]{
  width=\textwidth,
  fonttitle=\bfseries,
  breakable,
  fonttitle=\bfseries\color{Brown},
  colframe=Melon,
  colback=Melon!10
  #1}

\newtcolorbox[use counter=mynote]
  {mynote}[1][]
  {title=Note~\thetcbcounter,
   width=0.45\textwidth,
   left=0pt,
   right=0pt,
   fonttitle=\bfseries,
   coltitle=black,
   colframe=lightgray,
   colback=white,
   #1
}

\definecolor{main}{HTML}{5989cf}    
\definecolor{sub}{HTML}{cde4ff}     

\newtcolorbox{boxB}{
    fontupper = \bf\color{main}\footnotesize, 
    boxrule = 0.5pt,
    colframe = main,
    rounded corners,
    arc = 5pt   
}

\newtcolorbox{boxD}{
    fontupper = \small, 
    colback = sub, 
    colframe = main, 
    boxrule = 0pt, 
    toprule = 2pt, 
    bottomrule = 2pt 
}

\newtcolorbox{boxH}{
    fontupper = \small, 
    colback = sub, 
    colframe = main, 
    boxrule = 0pt, 
    leftrule = 6pt 
}

\newtcolorbox{boxG}{
    enhanced,
    boxrule = 0pt,
    colback = sub,
    borderline west = {1pt}{0pt}{main}, 
    borderline west = {0.75pt}{2pt}{main}, 
    borderline east = {1pt}{0pt}{main}, 
    borderline east = {0.75pt}{2pt}{main}
}    

\newtcolorbox{boxK}{
    fontupper = \small,
    sharpish corners, 
    boxrule = 0pt,
    toprule = 1.0pt, 
    enhanced,
    fuzzy shadow = {0pt}{-2pt}{-0.5pt}{0.5pt}{black!35} 
}

\newcommand{\secref}[1]{Sec.~\ref{#1}\xspace}
\newcommand{\figref}[1]{Fig.~\ref{#1}\xspace}
\newcommand{\tabref}[1]{Table~\ref{#1}\xspace}

\newcommand*\circled[1]{\tikz[baseline=(char.base)]{
            \node[shape=circle,draw,inner sep=0.5pt] (char) {#1};}}

\newcommand{\ie}{\textit{i.e.,}\xspace}
\newcommand{\eg}{\textit{e.g.,}\xspace}

\begin{document}
\bstctlcite{IEEEexample:BSTcontrol}
\title{

Measuring Emergent Capabilities of LLMs for Software Engineering: How Far Are We?

}

\author{\IEEEauthorblockN{Conor O'Brien, Daniel Rodriguez-Cardenas, Alejandro Velasco, David N. Palacio, and
Denys Poshyvanyk}
\IEEEauthorblockA{Department of Computer Science,
William \& Mary\\
Williamsburg, VA\\
\{csobrien, dhrodriguezcar, svelascodimate, danaderpalacio, dposhyvanyk\}@wm.edu}}


\maketitle

\begin{abstract}
The adoption of Large Language Models (LLMs) across multiple contexts has sparked interest in understanding how scaling model size might lead to behavioral changes, as LLMs can exhibit behaviors not observed in their smaller counterparts. Understanding these emergent capabilities is essential for advancing LLM development and improving their interpretability across diverse tasks. However, whether LLMs exhibit true emergence in the context of Software Engineering remains an unexplored topic, as most research has focused on NLP tasks. In this paper, we investigate the emergence of capabilities in the context of SE. We propose a model-agnostic pipeline for evaluating this phenomenon across three SE tasks: bug fixing, code translation, and commit message generation. More precisely, for each task, we present a case study instantiating our pipeline to analyze the emergence of capabilities in CodeGen1-multi across four scales ranging from 350M to 16.1B parameters. Our findings do not not provide evidence to support the idea of emergent capabilities resulting from scaling the model size in the selected set of tasks. We hope our results can pave the way to a more nuanced understanding of emergent capabilities of LLMs within the SE domain, guiding future research to focus on task-specific evaluations and the identification of alternative factors contributing to this phenomenon. Our work underscores the importance of task diversity in examining model behaviors and highlights potential limitations in transferring prior understandings of and approaches to emergence from NLP to Software Engineering.
\end{abstract}


\section{Introduction}

In the context of LLM evaluation,
emergent capabilities refer to abilities that large language models (LLMs) develop only when they reach a sufficiently large scale, such as increased training compute or parameter count, without showing gradual performance improvements beforehand. While performance often increases smoothly with model scale for certain tasks, researchers have observed tasks where LLMs exceed baseline performance only after reaching a specific scale \cite{wei2022emergent}. 


Being able to accurately measure emergent capabilities would enable LLM developers to make intelligent decisions regarding model scale and performance on certain tasks and how to train the models\cite{mitchell_measuring_2022}. LLMs are data centric models that improve as they grow and are able to observer larger data for training. Optimizing and identifying the sufficient data for training might reduce costs and improve the model performance.  For example, deciding how much compute to invest in a model architecture depends heavily on if there are plausible gains from doing so. If emergent capabilities are a prominent feature of LLMs in software engineering, then the choice to add scale to models makes sense even without predictable indications beforehand. Furthermore, since emergent capabilities may only be enabled by particular prompting strategies \cite{wei2022emergent}, it also suggests that researchers may try alternate prompting strategies at higher scales. Conversely, if emergent capabilities are not a prominent feature of LLMs, it suggests that if model performance does not increase beyond a certain scale, there would be little reason to continue making the model's scale larger.

Despite research efforts, the measurement of emergent capabilities in LLMs remains uncertain, particularly in the context of software engineering (SE). Most studies rely on visual inspection of scaling curves to demonstrate emergence \cite{wei2022emergent}, a method that is limited when only a few model scales are available. SE-focused LLMs, such as the CodeGen family \cite{nijkamp2023codegen,nijkamp2023codegen2}, lack the necessary range of model variants to support this approach. Furthermore, existing research has primarily concentrated on natural language processing tasks \cite{wei2022emergent,schaeffer2023emergent,lu2023emergent}, while code-related tasks have often been restricted to simplified examples \cite{srivastava2023imitation}. As a result, the evaluation of emergent capabilities in SE remains both poorly defined and largerly underexplored.



Consider the task of developing an LLM for automatic \textit{bug fixing}, starting with a small model and incrementally scaling its size. By their nature, emergent capabilities manifest only after a certain model scale is achieved in the form of rapid, unpredictable increases in performance. Understanding emergent capabilities in terms of scalability allows us to better understand the relationship between model scale and performance across this family of models. Hence, a precise and quantitative framework is needed. While one might hypothesize the general trend, a \textit{systematic method for evaluating emergent capabilities} would provide a clearer and more definitive assessment of how performance evolves with scale.

Given these limitations, this paper introduces \EmergentEval, a novel framework designed to measure and detect the emergence of LLM capabilities on tasks within the software engineering domain. 
We apply \EmergentEval to the \codegen family of models, evaluating emergent capabilities across three well-known SE tasks: \codeR, \ct, and \cm.
\EmergentEval determines whether a model family exhibits emergent capabilities (that is, capabilities the model gains unexpectedly after attaining large enough model scale)
based on the discontinuous Exact Match metric, the continuous BLEU and CodeBLEU metrics (for \codeR and \ct), and the continuous B-Norm and B-Moses metrics (for \cm)
and test cases from CodeXGLUE (for \codeR and \ct) and CoDiSum (for \cm)
The pipeline compares metric outcomes across model scales and applies regression analysis to identify unexpected patterns. The regression analysis serves to model our expectations of how the model's performance should be correlated with its scale in usual circumstances, and deviating far enough from this model of our expectations serves as a proxy variable for unpredictable, emergent growth.
In our case studies, we used \textit{linear regression}, as we expect the model's performance to increase linearly with its scale over the family of models we studied.


In our case studies, we evaluated the \codegen family of models ($350M$, $2.7B$, $6.1B$, and $16.1B$ parameters) on the selected SE tasks. We used metrics such as Exact Match, BLEU, and CodeBLEU, along with perturbation resistance measured using Ribeiro et al.'s Checklist \cite{ribeiro2020accuracy}. Perturbation resistance was analyzed by comparing the Levenshtein distance between unperturbed and perturbed inputs against performance changes. The obtained results present no evidence of emergent capabilities across the tested model scales.

This paper makes three key contributions. First, we introduce \EmergentEval, a framework for measuring and detecting model emergence through black-box evaluations, applicable to different SE tasks and broader research. Second, we evaluate the \codegen family of models on three SE tasks, finding no evidence of emergent capabilities. Lastly, a replication package including the source code and datasets used in our case studies \footnote{\url{https://github.com/WM-SEMERU/emergent-capabilities}}. 


The structure of this paper is as follows: In Section II we discuss the background related to measuring emergent capabilities on LLMs. In Section III, we discuss the definition of emergence we use in the context of software engineering, what metrics we use to evaluate the model's task performance, and how we measure emergence using those performances via \EmergentEval. In Section IV, we outline our research questions and the case studies we designed to answer each research question. In Section V, we showcase the results of our case studies. In Section VI, we discuss those results. In Section VII, we describe the potential threats to the validity of our studies. In Section VIII, we discuss the related work to this paper. in Section IX, we conclude and remark on the outlook of emergent capabilities in software engineering.

\section{Emergence in Software Engineering Background}\label{sec:background}

\subsection{LLMs for Code Generation}


Broadly speaking, LLMs are next-token predictors. They take as input a string of tokens (which can be considered words in its own language), and try to predict which token is most likely to follow the given tokens. For example, an LLM trained on Python code might see \texttt{for i in range} as context, and predict that the token \texttt{(} is likely to follow.

Most LLMs are advanced enough that, when given a textual description of a task, they infer that the most likely tokens to follow are the resolution of that task. This technique of providing targeted guidance as context to achieve a desired result is called \textit{prompting}. We might prompt that same LLM trained on Python code with \texttt{Add a comment to this code: print(lst[::-1])}, and it might infer that the most likely tokens to follow this prompt should be a comment, something like \texttt{\# prints the reversed version of lst}. We may prompt the model with this same template, providing it with a variety of code lines, and obtain a code-commenting prompt.

\subsection{Metrics in Software Engineering}

The metrics used in software engineering for evaluating LLM output are defined as comparing how similar two strings are, namely, the LLM generated output and the reference solution. Such metrics can be categorized into two categories: discontinuous  and continuous metrics. Similar to discontinuous and continuous functions in mathematics, a discontinuous metric is one where small changes to one of the two strings being compared may not be reflected in a similar change to the metric's result. Conversely, a continuous metric is one where such small changes are represented by corresponding small changes in the metric. 

To evaluate the quality of the outputs the model gives on each of the software engineering-related tasks, we use the Exact Match, BLEU \cite{Reiter2018BLEUValidity} (along with specializations B-Moses \cite{lin-och-2004-orange} and B-Norm \cite{lin-och-2004-orange}), and CodeBLEU \cite{ren2020codebleu} metrics.

Exact Match (\exm) is a discontinuous metric that simply computes the proportion of answers the model gives when completing a task that exactly matches the reference answer. \exm grades the model's responses with a score from 0 to 1, where higher is better. For example, if the model produces exactly the reference solution in 37 of the 100 given cases, the EM metric over the test cases is given as $37/100=0.37$. This metric is discontinuous because the resulting grade does not change as the model's answers change. Suppose the reference answer for one test case is \texttt{orange}; whether the answer the model gives is \texttt{asdf} or \texttt{orang}, the resulting grade does not change so long as the answer is not exactly what we expected.

\bleu is a metric designed to simulate human evaluation of machine-translated text \cite{papineni2002bleu}, which has since seen widespread use in NLP and ML \cite{Reiter2018BLEUValidity}.
The metric computes the proportion of $N$-grams (\ie runs of $N$ consecutive words) that appear in the model's answer against the number of $N$-grams that appear in the reference answer.
We adapt Lu et al.'s \cite{lu2021codexglue} implementation of the \bleu metric to support the variant metrics of B-Moses and B-Norm.
The B-Norm metric converts its inputs to lowercase text before grading, and both \bleu and B-Norm apply Lin et al. 2004 smoothing \cite{lin-och-2004-orange}; B-Moses neither applies smoothing nor converts its inputs to lowercase text.
The \bleu family of metrics grades the model on a score from 0 to 1, where higher is better.
These metrics are considered continuous since changing one string to be slightly more similar to the other results in a slightly better score.

\codebleu is a metric devised by Ren et al. \cite{ren2020codebleu} which adapts BLEU to process computer code rather than natural language text by considering the Abstract Syntax Tree (AST) structure of the code of the model's answer and the reference code. We use the implementation \verb|codebleu==0.6.1| hosted on Pypi \cite{codebleupypi2024}, which in turn is based on the aforementioned paper. Like BLEU, CodeBLEU grades the model on a score from 0 to 1, where higher is better.

To evaluate the model's ability to adapt to perturbed prompts, we analyze the relationship between the degree of perturbations and the resulting change in metric. Specifically, we use the Levenshtein distance metric to assess the degree of the perturbation, which counts the minimal number of certain character edits (via adding characters, removing them, or replacing one character with another) to transform the original prompt to the perturbed prompt. We expect that if a model truly has an emergent capability, that ability should be fairly robust and resist perturbation. Thus, with higher Levenshtein distance, the change in performance should be minimal; if model performance degrades with more perturbations (higher Levenshtein distance), then this would suggest the model is not adapting and may lack a truly general understanding of the task.

\subsection{Definition of Emergence}

Emergent capabilities in the context of families of LLMs which vary by scale are defined as those abilities the model has on a task under a metric with a certain prompting strategy under members of the family of a large enough scale, and not members of the family of lower scales. This is the so-called emergent scale \cite{wei2022emergent}. We note that model scale can be measured in terms of such factors as training compute and parameter size~\cite{wei2022emergent}.

Broadly, a model-task-metric-prompt quadruplet exhibits emergence if and only if the model performs poorly on the task at lower scales,
well on higher scales, and the improvement in performance does not linearly correspond with the increased scale;
in other words, emergent performance is characterized by unexpected and unpredictable jumps in performance. We define emergence to be a property of (model, task, metric, prompt) quadruplets.
A model may perform non-emergently on a variety of tasks, so it is important to distinguish between them when discussing emergence.
As noted by Wei et al. \cite{wei2022emergent}, it is possible for only a particular prompting strategy to perform emergently, even if other prompting strategies have predictable, non-emergent performance gains.
Keeping in mind that emergence can manifest only under specific metrics \cite{schaeffer2023emergent}, we include the researcher's choice of metric used to evaluate the model's performance into our definition of emergence
By exploring this emergency and the experience with these metrics, we aim to evaluate also the limitations of using only a handful of measurements for LLMs \cite{mitchell_measuring_2022}.

To determine whether a model exhibits emergence capabilities,
we need a set of models that only differ in size or scale.
This means emergence is a property of certain model sizes.
In theory, new models could be created at larger, smaller, or in-between sizes outside the ones we studied.
As a result, our method can only classify emergence based on the specific sizes we analyzed, not all possible sizes in the model family.



\section{Evaluating Emergence in Software Engineering}\label{sec:approach}

In this section, we introduce the methodology of our framework, \EmergentEval. We begin by providing a high-level definition of emergence in the context of SE, which can be instantiated in multiple ways. Next, we provide a comprehensive overview of the SE metrics used by our framework. Finally, we describe our approach and detail the emergence evaluation pipeline. 






\begin{figure}
    \centering
    \includegraphics[width=\linewidth]{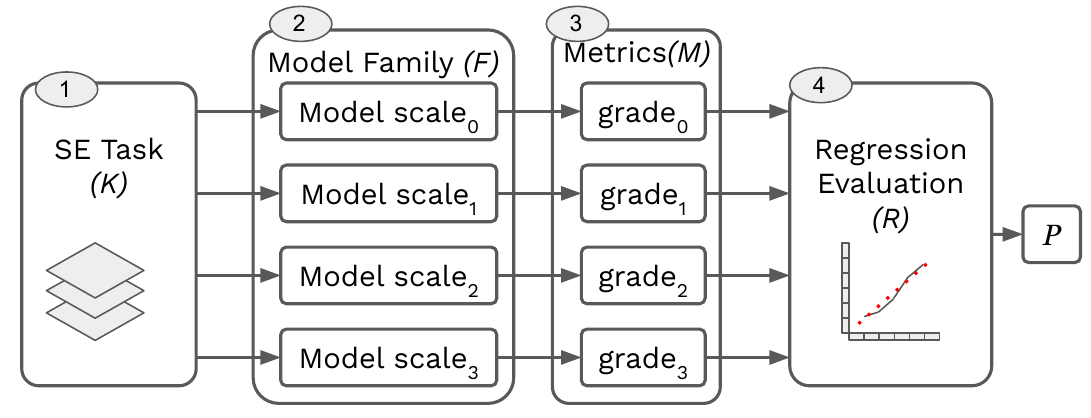}
    \caption{The \EmergentEval pipeline instantiated with particular test cases, four model scales, and a visualized linear regression evaluation, which produces a regression value $p$.}
    \label{fig:pipelineSolution}
\end{figure}

We designed \EmergentEval to evaluate whether a model’s ability to perform a given task exhibits emergent behavior. The framework’s pipeline requires the following inputs depicted at \figref{fig:pipelineSolution}: \circled{1} a task $K$, \circled{2} a model family $F$, \circled{3}  one or more metrics $M$, \circled{4} a regression method $R$ (producing a regression value $p$), and a regression threshold $t$. In this section, we will explain each of these components in detail and provide recommendations for effectively applying our framework. 

The regression $R$ is used to assess whether the model's results grow predictably. When applied to the individual scales $x$ from the model family $F_x$ the observed performances at each scale $y$, $R$ models the predicted values $\hat{y}$; the deviation is measured into a single value estimator, producing the regression value $p$. The regression value $p$ serves a similar function as the $p$-value in statistics, and the regression threshold $t$ similar to the level of significance value $\alpha$ in statistics; $t$ must similarly be chosen in advance. If $p\leq t$, \EmergentEval{} indicates the observed capabilities are within the predicted growth curve. Otherwise, if $p>t$, the pipeline indicates the observed capabilities are outside of the predicted growth curve and are emergent.

In our case studies, we opt to use a linear regression model. For evaluating linear regressions in the context of emergent capabilities, we recommend producing linear regression values $p$ with both Root Mean Square Deviation (RMSD) (Eq. \ref{eq:rmsd}) and Mean Absolute Error (MAE) (Eq. \ref{eq:mae}). 

\begin{equation}
\label{eq:rmsd}
\mathrm{RMSD}(x,y) = \frac{1}{n}\sqrt{\sum_{i=1}^n (y_i-\hat{y}_i)^2}
\end{equation}

\begin{equation}
\label{eq:mae}
\mathrm{MAE}(x,y) = \frac{1}{n}\sum_{i=1}^n \left|y_i-\hat{y}_i\right|
\end{equation}

To showcase \EmergentEval, we will consider one task from our first case study, \codeR, which we go into more detail on in the next section.

\circled{1} Our software engineering task $K$ consists of 100 test cases selected from the CodeXGLU Code repair dataset \cite{lu2021codexglue}. The task is to transform a given code fragment, \texttt{buggy\_code}, into its bug-free counterpart. For example, suppose \texttt{buggy\_code} is \texttt{public void METHOD\_1 ( TYPE\_1 VAR\_1 ) \{ this . VAR\_1 = VAR\_1 ; \} }, and we expect the repaired code to look like \texttt{public void METHOD\_1 ( TYPE\_1 VAR\_1 ) \{ \} }.

\circled{2} Our model family $F$ is a collection of the four models composing the \codegen model, with scales $350M$, $2.7B$, $6.1B$, and $16.1B$. We prompt each model to complete each test case in the task $K$, and record its answer. The answer the $16.1B$ model gives for the previous example is \texttt{public void METHOD\_1 ( TYPE\_1 VAR\_1 ) \{ \} }, as expected.

\circled{3} Let's choose our metric $M$ as just the Exact Match metric. For each test case in the task $K$, we compare the answer the model gave against the reference solution using our choice of metric. This results in an aggregate grade for how well the model performed on the task $K$ overall. In the case of \exm, the resultant grade is the number of answers the model gave which are exactly the reference solution, divided by the number of total test cases in the task (100). Applying this process to each of the four model scales, we obtain four corresponding grades: $0$, $0$, $0$, and $0.01$.

\circled{4} We use a linear regression evaluation for $R$. We will focus on using MAE (Eq. \ref{eq:mae}) to produce a regression value $p$, and consider a regression threshold $t=0.10$. First, we construct a linear model using a least-squares linear regression on our model scales $x=[0.35, 2.7, 6.1, 16.1]$ and our observed grades $y=[0, 0, 0, 0.01]$, obtaining $\hat{y}\approx -0.018+0.007x=[-0.015,  0.001, 0.024, 0.091]$. We then evaluate how well this model fits our data by calculating $\frac{1}{4}\sum_{i=1}^{4}\left|y_i-\hat{y}_i\right|=(|0-(-0.015)|+|0-0.001|+|0-0.024|+|0.01-0.091|)/4=(0.015+0.001+0.024+0.081)/4=0.032$. This gives us $p=0.032$, which we can then compare against our threshold $t=0.10$. As $p\ll t$, we can conclude the model's performance gains are linear and non-emergent.


When applying \EmergentEval to assessing LLMs, we recommend considering two primary cases: First, applying the pipeline to the selected tasks in the base case. Second, in the perturbed case, generating a new task using Ribeiro et al.'s \cite{ribeiro2020accuracy} CheckList utility  which are perturbations on the original task. By assessing the models from multiple angles, we will be able to assert whether the model exhibits emergent capabilities more confidently.

\section{Case Study Design}


We proposed three research questions and designed corresponding case studies to demonstrate the use of \EmergentEval{}, evaluate its effectiveness in each scenario, and determine whether the model's performance in these scenarios exhibits emergent capabilities. 

\subsection{Experimental Context}

\textit{Models:} For our analysis, we examined the \codegen family of models \cite{nijkamp2023codegen}, which are available in parameter counts of $350M$, $2.7B$, $6.1B$, and $16.1B$. We applied a regression threshold of $t=0.10$ for both RMSD and MAE. This threshold was established before testing by generating example data representing smooth and emergent curves and selecting a $t$-value that approximately separates the two categories.

\begin{enumerate}[label=\textbf{RQ$_{\arabic*}:$}, ref=\textbf{RQ$_{\arabic*}$}, wide, labelindent=5pt]\setlength{\itemsep}{0.2em}
  \item \label{rq:codepair} \textit{How likely does \textbf{\codeR} emerge as a software engineering capability in an LLM for code?} 
  \item \label{rq:codetraslation} \textit{How likely does \textbf{\ct} emerge as a software engineering capability in an LLM for code?}
  \item \label{rq:commit}  \textit{How likely does \textbf{\cm} emerge as a software engineering capability in an LLM for code?}
\end{enumerate}

\textit{Testbeds:} We sampled subsets dataset form CodeXGLUE\cite{lu2021codexglue} for \codeR and \ct. Particularly, we extracted the first 100 cases from Code-Code/code-refinement for building the \codeR. To generate the \ct testbed we used the first 100 cases from Code-Code/code-to-code-trans. Finally, to generate \cm testbed, we examined datasets from CoDiSum \cite{ijcai2019p552} and Zhang et al. \cite{zhang2024using}, and decided to test on the CoDiSum dataset (first 100 cases).

\textit{Prompting strategy:} As emergent capabilities are a feature of prompting strategy \cite{wei2022emergent}, we configured a prompt strategy per task. Successfully interacting with the CodeGen1-multi model with prompting requires formatting the prompts in the framework of code completion, most typically through code comments which describe the purpose of the desired code to generate. The more conventional prompting strategies structured as natural language proved extremely ineffective during testing.

\subsection{\ref{rq:codepair} Code Repair Methodology}

\begin{figure}[h]
    \centering
    \includegraphics[width=0.9\linewidth]{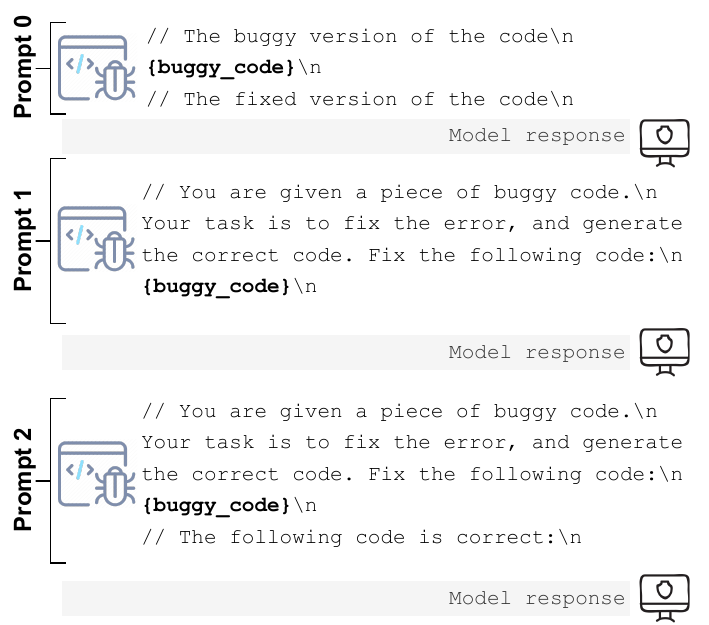}
    \caption{Prompting strategy used on CodeGen1-multi for the code repair tasks.}
    \label{fig:code-repair-prompts}
\end{figure}

To answer \ref{rq:codepair} and determine the effect varying scale has upon model performance in light of our definition of emergence and \EmergentEval{}, we must design a variety of prompting strategies (see Fig. \ref{fig:code-repair-prompts}), as different prompting strategies can enable the model to acquire emergent capabilities. We must then apply each prompt to the \EmergentEval{} pipeline with the four CodeGen1-multi models, grade each's performance on the CodeXGLUE test cases, and assess each according to our linear regression models under RMSD and MAE.

Then, we employ bootstrapping to resample our data and investigate the underlying structure beyond the mere aggregate statistics. Bootstrapping works by taking a random subsample of $S$ of the model's answers, and using each of our metrics on this subsample. We repeat this $N$ times to get a cloud of points for each model scale in the family. We configure our bootstrapping with a subsample size $S=50$ over $N=500$ iterations~\cite{obrien2024emergentse}.

\subsection{\ref{rq:codetraslation} Code Translation Methodology}

\begin{figure}
    \centering
    \includegraphics[width=0.9\linewidth]{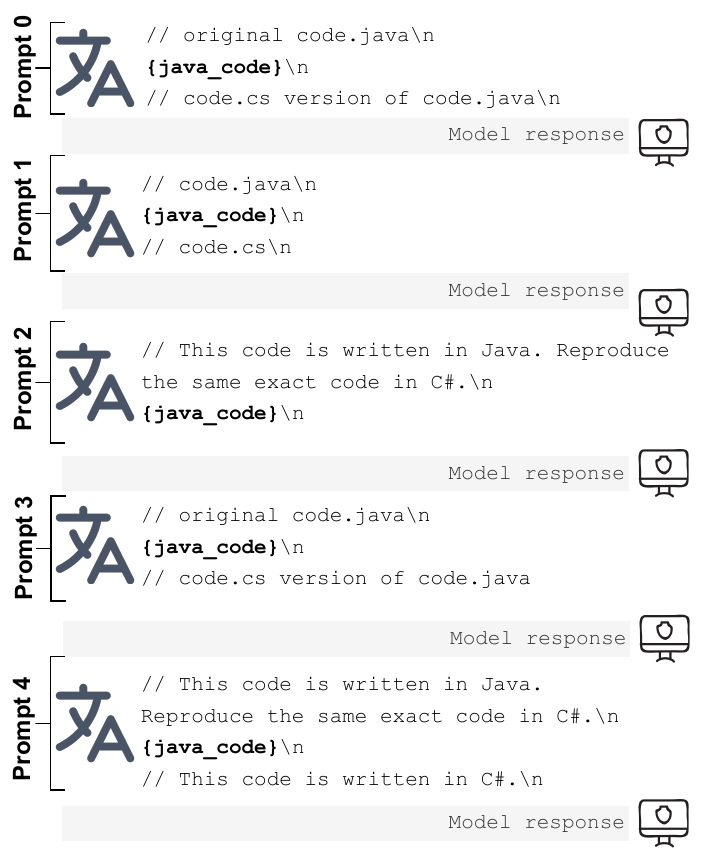}
    \caption{Prompting strategy used on CodeGen1-multi for the code translation tasks.}
    \label{fig:code-translation-prompts}
\end{figure}

Similarly to answering \ref{rq:codepair}, we answer \ref{rq:codetraslation} by designing a variety of prompts (see Fig. \ref{fig:code-translation-prompts}), applied to our \EmergentEval{} pipeline, and utilizing bootstrapping with subsample size $S=50$ and $N=500$.

\subsection{\ref{rq:commit} Commit Message Generation Methodology}

\begin{figure}[h]
    \centering
    \includegraphics[width=0.9\linewidth]{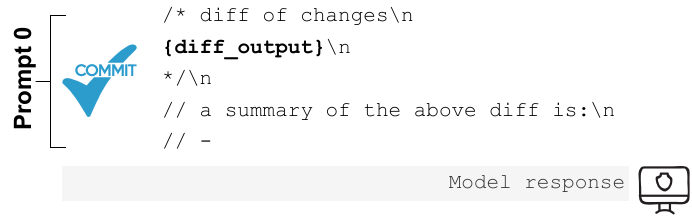}
    \caption{Prompting strategy used on CodeGen1-multi for the commit message generation tasks. When designing this prompt, coaxing the model to output English description required finesse. In this case, during testing, we found the most consistent approach to obtaining English descriptions was to embed the \texttt{diff} output in a labeled, multi-line code comment, followed by two single-line comments which prompt the model with the context of English language descriptions contained within code comments.}
    \label{fig:commit-message-generation-prompts}
\end{figure}

Due to the nature of this task intrinsically exhausting the model architecture's resources, we designed only one initial prompt (see Fig. \ref{fig:commit-message-generation-prompts}) and applied it to our \EmergentEval{} pipeline; we did not employ bootstrapping here, as the resultant grades were 0 almost universally.

\subsection{\EmergentEval Exploratory Analysis}

To evaluate the code repair task under \ref{rq:codepair} (automatic bug fixing), we tested the models as described at \secref{sec:approach} on selected tasks from the \codeR dataset. Each \codeR task consists of Java methods consolidated onto a single line, with variable names, types, strings, and other identifiers consistently replaced by generic placeholders such as \verb|TYPE_1| and \verb|STRING_3|. The objective is to generate the corrected version of the given code snippet. We assessed model performance using the \exm, \bleu, and \codebleu metrics, comparing the model's generated output against the expected output. We configure \bleu and its metric specializations, B-Norm and B-Moses, with $N$-grams sized up to $N=4$.

Second, to evaluate the code translation task under \ref{rq:codetraslation}, we tested the models on CodeXGLUE's \cite{lu2021codexglue} CodeTrans task (also known as code-to-code translation). The CodeTrans task involves translating Java methods, consolidated onto a single line, into their equivalent C$^\sharp$ code. Similar to the bug fixing task, we assessed model performance using the \exm, \bleu, and \codebleu metrics, comparing the model's generated output against the expected output.

Lastly, to evaluate the commit message generation task under \ref{rq:commit}, we tested the models on CoDiSum's dataset \cite{ijcai2019p552} using BLEU. Each task provides the output of the \texttt{diff} command, which highlights changes across one or more files, and expects an English description summarizing these changes. We assessed model performance using a subset of the metrics applied by Zhang et al. \cite{zhang2024using}, specifically B-Moses and B-Norm, as described in detail in Tao et al.'s study on commit message generation models~\cite{tao2021evaluation}.


\section{Results}\label{sec:results}


\subsection{\ref{rq:codepair}: Bug Fixing}

\begin{figure}
    \centering
    \includegraphics[width=0.9\linewidth]{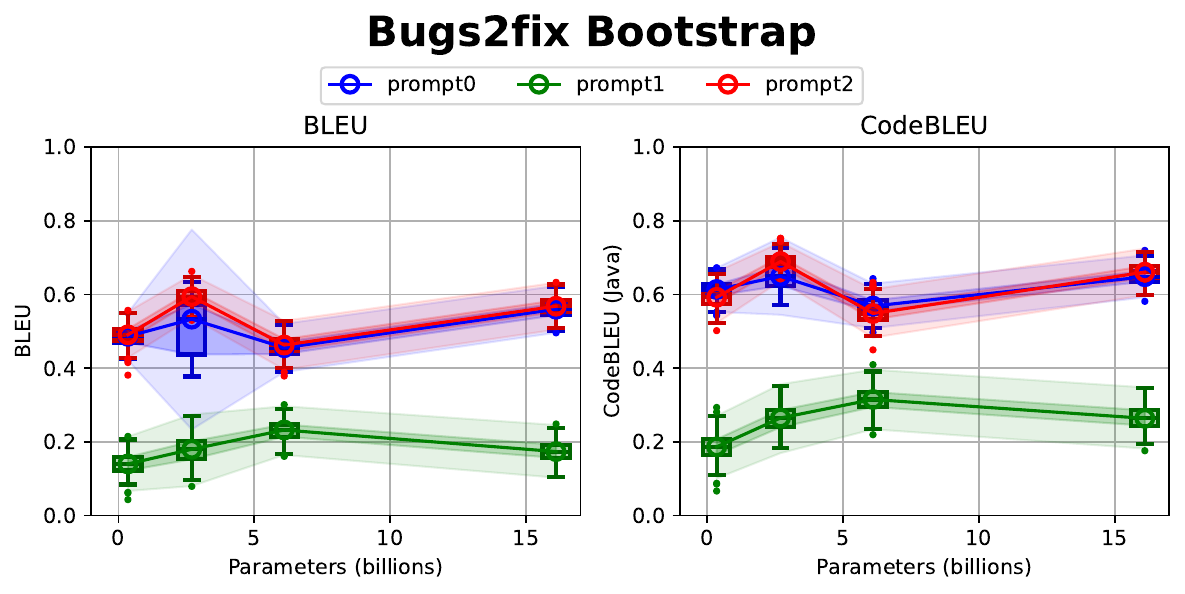}
    \caption{Bootstrapping with $S=50$ and $N=500$ over the CodeXGLUE Bugs2fix task, assessed by metrics BLEU and CodeBLEU, over the results of prompting CodeGen1-multi. (The Exact Match metric is omitted, as it is 0 across all data points).}
    \label{fig:ch4:b2f-bootstrap}
\end{figure}

\figref{fig:ch4:b2f-bootstrap} depicts the results when evaluating \codegen with proposed prompts at \figref{fig:code-repair-prompts}. The exact match metric highlights that the models are not producing the exact expected output, which is not necessarily a negative outcome. This variation indicates that the models are not overly biased or simply parroting the expected responses. When evaluating proposed prompts on the code repair tasks, we observe that the model achieved middling performance under the \bleu and \codebleu metrics, and very low performance under the \exm metric (see \figref{fig:ch4:b2f-bootstrap}). Furthermore, the results were decisively low (see Table \ref{tab:results-b2f}), with all calculated regression values $p$ much less than the regression threshold $t=0.10$.

\begin{figure}
    \centering
    \includegraphics[width=0.9\linewidth]{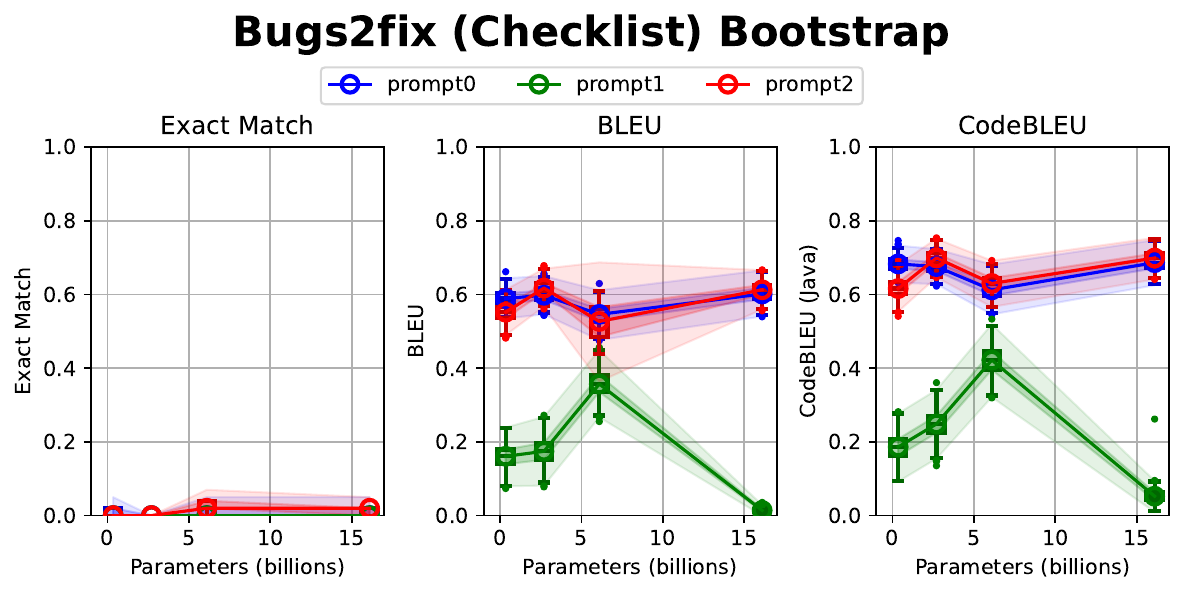}
    \caption{Bootstrapping with $S=50$ and $N=500$ over the CodeXGLUE Bugs2fix Checklist perturbed task, assessed by metrics Exact Match, BLEU, and CodeBLEU, over the results of prompting CodeGen1-multi.}
    \label{fig:ch4:b2f-cl-bootstrap}
\end{figure}


\begin{table}[]
\centering
\begin{tabular}{llll}
\hline
\textbf{Metric}          & \textbf{Prompt}  & \textbf{RMSD}   &  \textbf{MAE}    \\ \hline
EM              & \prom & 0      & 0      \\
                & \promi & 0      & 0      \\
                & \promii & 0.0015 & 0.0012 \\ \hline
BLEU            & \prom & 0.0252 & 0.0218 \\
                & \promi & 0.0307 & 0.0256 \\
                & \promii & 0.0526 & 0.0449 \\ \hline
CodeBLEU (Java) & \prom & 0.0304 & 0.0248 \\
                & \promi & 0.0412 & 0.0375 \\
                & \promii & 0.0537 & 0.0453 \\ \hline
\end{tabular}
\caption{Results for evaluating \EmergentEval on CodeGen1-multi for code repair on each metric. All values are in the range $[0,0.06)$, far below the regression threshold of $t=0.10$~\cite{obrien2024emergentse}.}
\label{tab:results-b2f}
\end{table}


As for the \textit{checklist} perturbations, when evaluating our three prompts on the perturbed code repair tasks, we found that the model achieved better performance under all three metrics (see \figref{fig:ch4:b2f-cl-bootstrap}).  \figref{fig:ch4:b2f-cl-bootstrap} depicts slight improvements when scaling the model at both metrics \bleu and \codebleu. However, this slight improvement does not represent an emergent capability. Furthermore, the results were overall low (see \tabref{tab:b2fcl-results}), with most calculated regression values $p$ much less than the regression threshold $t=0.10$. Two regression values for \texttt{prompt1} under the RMSD formula produced regression values $p>0.10$, although not because of emergent growth, but because of a sharp decline in performance. This may be considered negative emergent growth, although as it only appears under one of the two formulae for producing $p$ values, we do not draw this conclusion.

Important to note is the marked decrease in performance under \texttt{prompt1} at the $16.1B$ parameter scale. Whereas the test cases were formatted as code snippets without line breaks, and the model was supposed to respond in kind, the $16.1B$-parameter CodeGen1-multi model instead emitted single-line comments. Further inspection of what the model would have predicted beyond the first line shows code formatted more conventionally, rather than all run together on a single line. Thus, this seems to be more likely a limitation of the testing harness setup we used clashing with the potentially negatively emergent capability of lack of following the syntactic context, rather than a negative semantic emergent capability. A more advanced testing harness might attempt to take into consideration ASTs as output, \eg requesting the model generate until it has provided a complete AST, or until it exceeds a certain limit.


\begin{table}[]
\centering
\begin{tabular}{llll}
\hline
\textbf{Metric}          & \textbf{Prompt}  & \textbf{RMSD}            & \textbf{MAE}    \\ \hline
EM              & \prom & 0.0041          & 0.0033 \\
                & \promi & 0               & 0      \\
                & \promii & 0.0073          & 0.0063 \\ \hline
BLEU            & \prom & 0.0231          & 0.0199 \\
                & \promi & \textbf{0.1037} & 0.0890 \\
                & \promii & 0.0326          & 0.0278 \\ \hline
CodeBLEU (Java) & \prom & 0.0294          & 0.0254 \\
                & \promi & \textbf{0.1145} & 0.0952 \\
                & \promii & 0.0312          & 0.0264 \\ \hline
\end{tabular}
\caption{Results for evaluating \EmergentEval{} on CodeGen1-multi on the perturbed test cases for code repair on each metric. Bolded are regression values $p$ which exceed the regression threshold value $t=0.10$, which occur under \texttt{prompt1} and RMSD; MAE metrics, while not exceeding the threshold, are close.for RMSD and MAE. It is important to note these are nonlinear due to extreme negative performance, rather than emergent growth. Other prompting strategies lie within the range $[0,0.04)$~\cite{obrien2024emergentse}.}
\label{tab:b2fcl-results}
\end{table}

\begin{figure}[h]
    \centering
    \includegraphics[width=0.9\linewidth]{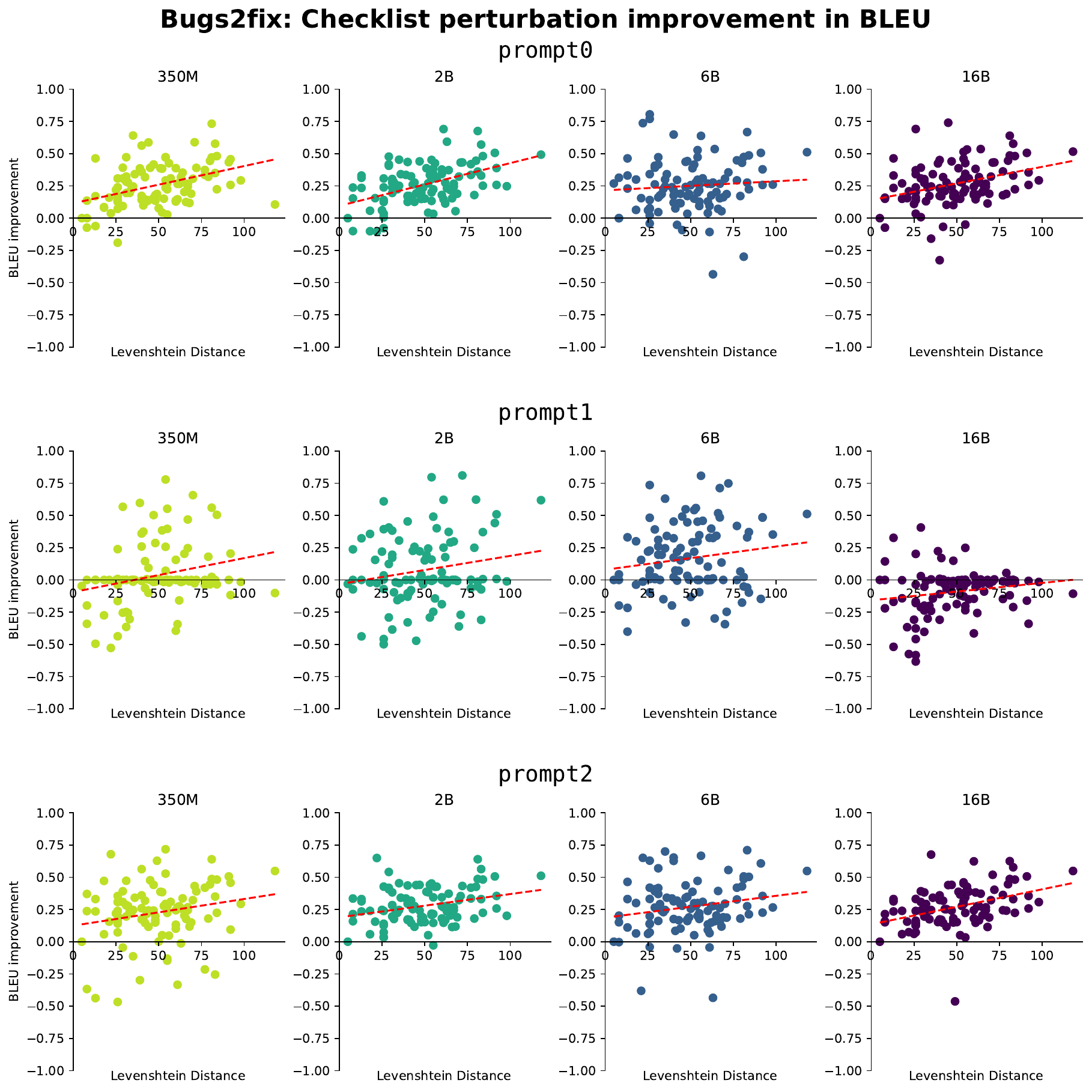}
    \caption{Scatterplot of the relationship between the Levenshtein distance between the original, unmodified test case and the corresponding perturbed test case on the x-axis, and the relative performance increase from the BLEU score of the unmodified test case to the BLEU score of the perturbed test case; positive y-values indicate improvement. }
    \label{fig:ch3b2fcl-sca1}
\end{figure}
\begin{figure}[h]
    \centering
    \includegraphics[width=0.9\linewidth]{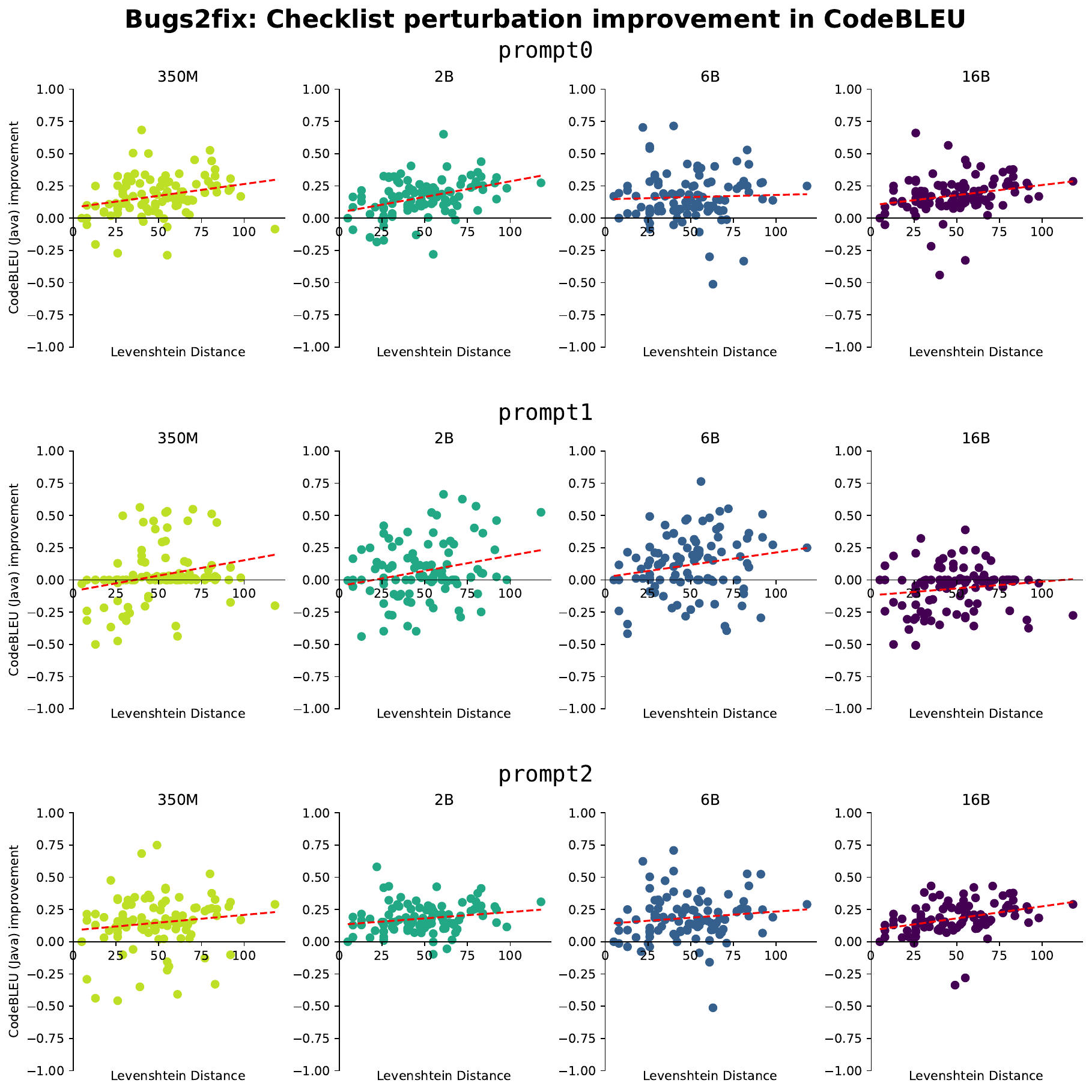}
    \caption{Scatterplot of the relationship between the Levenshtein distance between the original, unmodified test case and the corresponding perturbed test case on the x-axis, and the relative performance increase from the CodeBLEU score of the unmodified test case to the CodeBLEU score of the perturbed test case; positive y-values indicate improvement~\cite{obrien2024emergentse}.}
    \label{fig:ch3b2fcl-sca2}
\end{figure}

We can model the relationship between the Levenshtein distance of the perturbation from the original, unperturbed prompt, and the resulting change in performance (see Fig. \ref{fig:ch3b2fcl-sca1} for BLEU and Fig. \ref{fig:ch3b2fcl-sca2} for CodeBLEU). Overall, this shows a very loose but generally positive correlation between perturbation and performance increase. We attribute this to the nature of the perturbations tending to resemble more natural code rather than the prompt's given placeholders (\eg replacing the constant \texttt{STRING\_0} in one task with an actual string, say \texttt{"orange"}).

\begin{boxK}
\ref{rq:codepair} \codeR: {On the observed scale over the dataset, no positive emergent capabilities were performed for the \codeR task, as almost all regression values had $p \ll 0.10$. A few regression values exceeded $p > 0.10$, but only under RMSD and in a negative direction.}
\end{boxK}


\subsection{\ref{rq:codetraslation}: Code Translation}

\begin{figure}[h]
    \centering
    \includegraphics[width=0.9\linewidth]{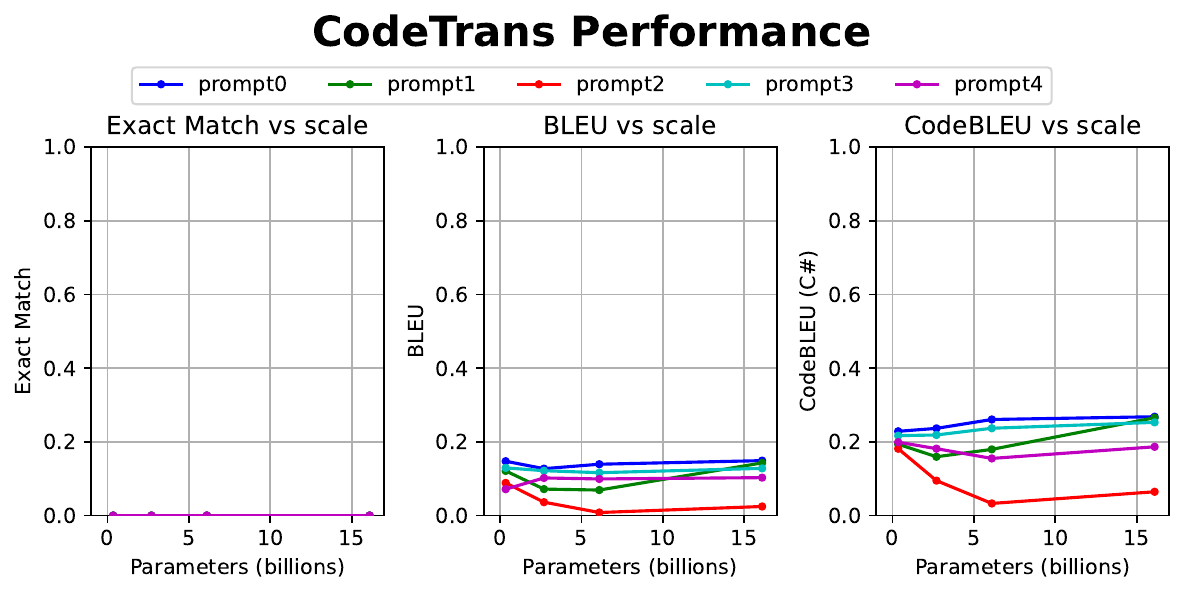}
    \caption{CodeGen1-multi performance on CodeXGLUE Code2code test cases, assessed by the metrics Exact Match, BLEU, and CodeBLEU (operating on C$^\sharp$ ASTs)~\cite{obrien2024emergentse}.}
    \label{fig:ch3c2call}
\end{figure}

When evaluating our five prompts on the code traslation task, we observed that the model consistently exhibited poor performance across all metrics. \figref{fig:ch3c2call} illustrates the plots for exact match, \bleu and \codebleu scores, corresponding to the four prompts detailed in \figref{fig:code-translation-prompts}. The plots display the scores for each \codegen model scale, revealing no significant improvements or jumps as the model scales. The \bleu metric consistently reports scores below 0.2, while \codebleu remains below 0.3 across all scales. Additionally,  \tabref{tab:results-c2c}) summarizes the RMSD and MAE values, with all calculated regression $p$ beign significantly lower than the threshold $t=0.10$.


\begin{table}[]
\centering
\begin{tabular}{llll}
\hline
\textbf{Metric}                & \textbf{Prompt}  & \textbf{RMSD}   & \textbf{MAE}    \\ \hline
BLEU                  & \prom & 0.0078 & 0.0063 \\
                      & \promi & 0.0269 & 0.0254 \\
                      & \promii & 0.0245 & 0.0227 \\
                      & \promiii & 0.0052 & 0.0046 \\
                      & \promiv & 0.0103 & 0.0092 \\ \hline
CodeBLEU (C$^\sharp$) & \prom & 0.0074 & 0.0064 \\
                      & \promi & 0.0201 & 0.0190 \\
                      & \promii & 0.0442 & 0.0401 \\
                      & \promiii & 0.0038 & 0.0031 \\
                      & \promiv & 0.0159 & 0.0129 \\ \hline
\end{tabular}
\caption{Results for evaluating \EmergentEval{} on CodeGen1-multi for code translation on each prompt and metric. We omit EM metrics, as they were 0 throughout. There is low variability, and few values occur outside the range $[0,0.03)$, most notably \texttt{prompt2} under CodeBLEU achieving RMSD and MAE $>0.40$. Still, all values are well below the regression threshold $t=0.10$~\cite{obrien2024emergentse}.}
\label{tab:results-c2c}
\end{table}


\begin{figure}
    \centering
    \includegraphics[width=0.9\linewidth]{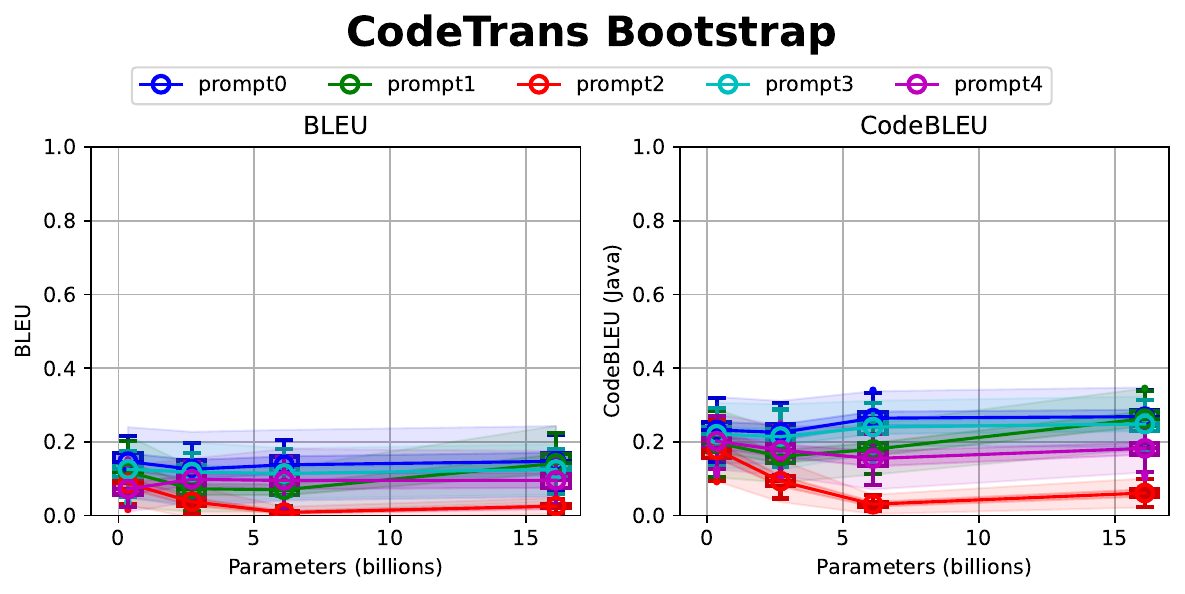}
    \caption{Bootstrapping with $S=50$ and $N=500$ over the CodeXGLUE CodeTrans task, assessed by metrics Exact Match, BLEU, and CodeBLEU, over the results of prompting CodeGen1-multi~\cite{obrien2024emergentse}.}
    \label{fig:ch4:c2c-bootstrap}
\end{figure}

We employed statistical bootstrapping to understand the spread and variation of our samples (see Fig. \ref{fig:ch4:c2c-bootstrap} for bootstrapping).

\begin{boxK}
\ref{rq:codetraslation} \ct: {On the observed scale over the dataset, we did not observe emergent capabilities for the Code Translation task, with regression values $p\ll 0.10$.}
\end{boxK}


\subsection{\ref{rq:commit}: Commit Message Generation}
\begin{figure}
    \centering
    \includegraphics[width=0.9\linewidth]{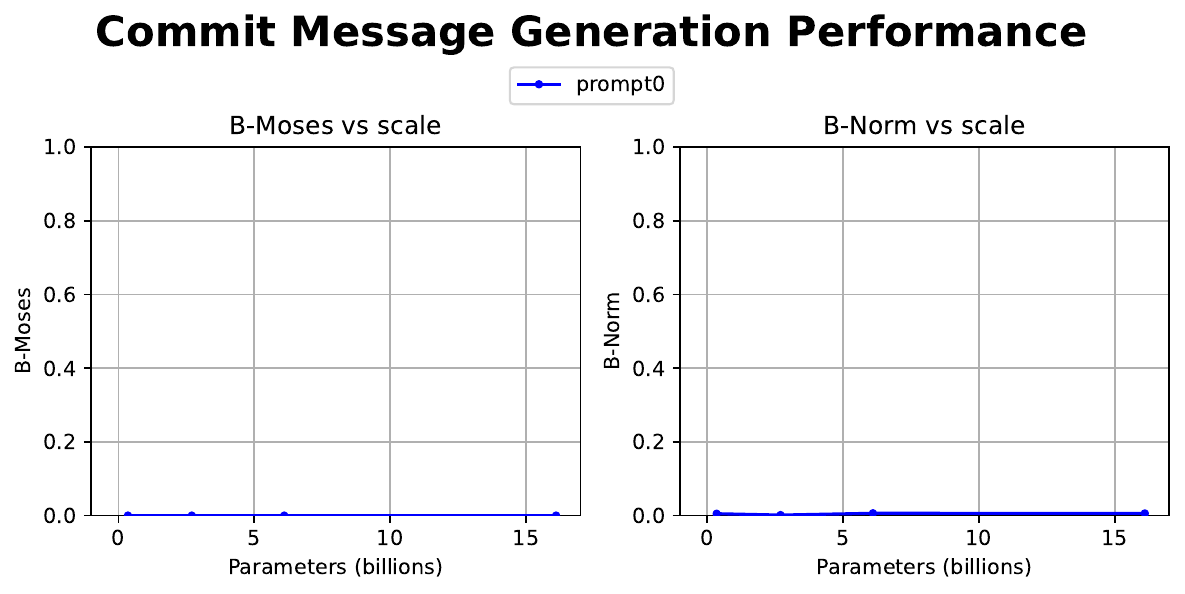}
    \caption{CodeGen1-multi performance on test cases from the CoDiSum dataset, assessed by the metrics B-Moses and B-Norm~\cite{obrien2024emergentse}.}
    \label{fig:ch3cmgall}
\end{figure}

When evaluating our prompts on the commit message generation task, we found that the model achieved poor performance under all metrics as presented at \figref{fig:ch3cmgall}). Furthermore, the results were decisively low, with all calculated regression values $p$ much less than the regression threshold $t=0.10$ (see Table \ref{tab:results-cmg}). Fitting the entire diff in the model's working memory proved infeasible for many prompts, leading to errors caused by insufficient resources within the model architecture.

\begin{table}[]
\centering

\begin{tabular}{llll}
\hline
\textbf{Metric}  & \textbf{Prompt}  & \textbf{RMSD}   & \textbf{MAE}    \\ \hline
B-Moses & \prom & 0      & 0      \\ \hline
B-Norm  & \prom & 0.0017 & 0.0015 \\ \hline
\end{tabular}
\caption{Results for evaluating \EmergentEval{} on CodeGen1-multi for commit message generation on each metric. Results are 0 or incredibly close to 0, signifying almost perfect linearity~\cite{obrien2024emergentse}.}
\label{tab:results-cmg}
\end{table}


\begin{boxK}
\ref{rq:commit} \cm: {On the observed scale over the dataset, no emergent capabilities were identified for the Commit Message Generation task, with regression values consistently at $p \ll 0.10$.
}
\end{boxK}

\section{Discussion}

We believe the obtained negative results in the case studies were obtained due to an apparent linearity in the obtained curves from the  regression. Nothing suggests that there are any significant unexpected jumps in model performance under any of our metrics for any of our tasks when comparing against the model scale. Moreover, the model performance roughly increases linearly with model scale.

Although increasing the model scale generally improves performance, we found that prompting techniques play a more decisive role in determining model performance. In fact, with certain prompts, increasing model scale can negatively impact performance. Interestingly, emergent capabilities were absent across all model scales and tasks, regardless of the prompting methods employed. This outcome might be attributed to the fact that Wei et al. (\cite{wei2022emergent}, Appendix D) observed emergence only in models with parameter counts ranging from tens of billions to a trillion. Another possibility is that our experiments did not identify the specific prompts needed to unlock emergent capabilities. Alternatively, the relative stability and predictability of the CodeGen1-multi model at these smaller scales may explain the lack of observed emergence.

The Checklist perturbed code repair task elicited nonlinear results under \texttt{prompt1} and RMSD (see Fig. \ref{fig:ch4:b2f-cl-bootstrap} and Table \ref{tab:b2fcl-results}) through the application of \EmergentEval{}, which seems to suggest \textit{negative} emergence. This conclusion comes with two observations which may somewhat limit the conclusion here. First, it is important to note that this result straddles the pipeline's verdict of emergence, producing a regression value $p$ slightly above $t=0.10$ under RMSD, and slightly below it under MAE. Second, the nature of the potential negative emergence is likely less reflective of the model's inability to reason about the code using this prompting strategy, and more of formatting the code similarly to what was given, as mentioned in results section\ref{sec:results}.



\section{Threats to Validity}

\textbf{Construct validity.} Through our testing, we determined that the primary cause for improved performance was not model scale, but rather prompting technique. This means that our testing framework is not sufficient to discover emergent capabilities based on  improvements intrinsic to the model. Furthermore, across our three scenarios, we tested only a handful of prompts. As emergence is thought to be a phenomenon which directly depends on choice of prompting technique, there may simply be more effective prompts for the CodeGen1-multi model in the given test cases.

\textbf{Internal validity.} The continuous metrics used in this study (BLEU, including B-Moses and B-Norm, and CodeBLEU) are more concerned with the model's outputs being \textit{apparently similar} to reference solutions, instead of them being \textit{correct} or \textit{useful}. Although Shaeffer et al. claim emergent capabilities disappear when using continuous metrics \cite{schaeffer2023emergent}, we acknowledge that these metrics may lose sight of our actual goals of model correctness and usefulness, a goal in which the discussion of emergent capabilities is ultimately grounded. Therefore, used metric migth be insufficient to measure and observe potential emergent capabilities. To deal with this limitation we also used levenshtein metric to observe how disperse is the outcome from the input as an indicator of emergent performance. 

\textbf{External validity.} Although EmergenceEval  generalizes well to other models and target domains other than software engineering, our results may generalize less well. The CodeGen1-multi model is somewhat unique among LLMs in that it is a specialist in software engineering tasks and is not receptive to conventional prompting techniques with more general LLMs, instead relying on arcane prompting via suitably contextualized code comments; the relationship between ``prompt'' and model behavior is much hazier with the CodeGen series of models.



\section{Related Work}

The notion of emergent capabilities has long been a subject of inquiry in the field of machine learning. As early as 2020 and 2021, researchers were concerned about the negative side effects of increasing model scale, particularly with increasingly larger models adopting the explicit and implicit biases featured in their training sets, and their ability to mimic coherent human speech and articulation \cite{bender2021parrots}, based on concerns and research done into the various kinds of bias found in models such as BERT, GPT-2, and GPT-3~\cite{Guo_2021,hutchinson-etal-2020-social}.

Researchers have used the term \textit{emergence} to refer to capabilities models acquired through training that they were not explicitly trained for, as in Nijkamp et al.'s paper documenting the CodeGen family, where the term is used to describe the model's capability to synthesize programs from comment descriptions \cite{nijkamp2023codegen}. The rigorous study of emergent capabilities, however, was made most prominent with Wei et al., where emergent capabilities are framed not as capabilities acquired without explicit intentions, but as sharp, unpredictable jumps in performance \cite{wei2022emergent}. Their findings suggest that LLMs, through certain tasks and prompting methods, when scaled high enough (\eg by training compute or parameter count), can unexpectedly and noticeably break plateaued performance.

However, various researchers contest the claim that these models exhibited emergent capabilities whatsoever. Shaeffer et al. \cite{schaeffer2023emergent} suggest the appearance of emergence is better explained by the metrics used to assess model performance, rather than as some property of the model itself; they implicate nonlinear and discontinuous metrics as a confounding factor in Wei et al.'s results. Other research by Lu et al. suggests what appears to be emergent capabilities are better explained as the results of in-context learning, that is, the model's ability to derive crucial information from the context it is prompted with and apply that knowledge to the task at hand~\cite{lu2023emergent}.

Following Schaeffer et al. \cite{schaeffer2023emergent}, the most important metrics to consider when evaluating model emergence are continuous metrics, as discontinuous metrics may induce the mirage of emergence. Although the exact nature of these tests can vary depending on the experimental setup, we choose to evaluate linear regressions on the attained data (model performance graded by various metrics) to assess for non-linear growth. Non-linear growth is a proxy variable for emergent growth; as we only have a limited view of the true shape of the performance graph as seen through four different model configurations, we choose to evaluate a linear regression model rather logarithmic or exponential regression models to avoid overfitting the curves to the limited datapoints. This restriction does not exist generally, and using \EmergentEval{} in circumstances where more data is available demands considering additional, more sophisticated regressions.


\section{Conclusions and Future Work}

We find that varying the model scale of the CodeGen1-multi model from $350M$ parameters to $16.1B$ parameters generally tends to improve performance slightly in a linear fashion, albeit somewhat loosely; performance was not usually strictly increasing, nor was maximal performance always attained with maximal tested model scale.

Being unable to identify emergent capabilities in this case gives credence to the general consensus that emergent capabilities may be an illusory phenomenon, even in the domain of software engineering. It implies that, when developing LLMs for software engineering, merely adding additional model scale (\eg compute or parameter count) is not justified without seeing incremental performance gains beyond a certain point, and that alternative solutions to increase model performance would be required in such cases.

Reasoning about emergent capabilities is a difficult task, fraught with doubts about the existence of the phenomenon itself, as well as uncertainties as to discovering the right prompts and the proper use of metrics. We hope that future research can utilize our \EmergentEval{} pipeline to aid future investigations into the question of emergent capabilities.

Whether models can exhibit emergent capabilities is a double-edged sword. The prospect of untapped potential awaiting model designers, if only they provide the requisite increased scale, is both tantalizing and trepidating, depending on the exact nature of the emergent capabilities gained. The model may gain, say, superlative reasoning abilities emergently; it may also gain superlative discriminatory abilities emergently. We saw the potential for negative emergence in the case study answering \ref{rq:codepair}. Conversely, identify no emergence on a given scale can also establish stability.

Examining the question of emergence is an important step into assessing the interpretability, stability, and predictability of a model. While most models are not released to be available by discrete model scales, those training models can assess the intermediate models scaled by training FLOPs for signs of emergence using our pipeline.

Future work regarding emergent capabilities in software engineering should focus on acquiring and testing models on a more granular scale so as to enable more sophisticated regressions which might more accurately represent and interpret the model's growth.

\bibliographystyle{IEEEtran}
\bibliography{utils/main}

\end{document}